\PassOptionsToPackage{dvipsnames}{xcolor}

\documentclass[10pt, conference]{IEEEtran}
\IEEEoverridecommandlockouts
\usepackage{cite}
\usepackage{graphicx}
\usepackage{xcolor}
\def\BibTeX{{\rm B\kern-.05em{\sc i\kern-.025em b}\kern-.08em
    T\kern-.1667em\lower.7ex\hbox{E}\kern-.125emX}}

\usepackage{listings}
\usepackage[dvipsnames]{xcolor}

\newcommand{\tool}{\textsc{CrySol}\xspace}

\usepackage[hidelinks]{hyperref}
\definecolor{verylightgray}{rgb}{.97,.97,.97}

\lstdefinelanguage{Solidity}{
  keywords=[1]{anonymous, assembly, assert, balance, break, call, callcode, case, catch, class, constant, continue, constructor, contract, debugger, default, delegatecall, delete, do, else, emit, event, experimental, export, external, false, finally, for, function, gas, if, implements, import, in, indexed, instanceof, interface, internal, is, length, library, log0, log1, log2, log3, log4, memory, modifier, new, payable, pragma, private, protected, public, pure, push, require, return, returns, revert, selfdestruct, send, solidity, storage, struct, suicide, super, switch, then, this, throw, transfer, true, try, typeof, using, view, virtual, while, with, addmod, ecrecover, abi, encode, encodePacked, keccak256, mulmod, ripemd160, sha256, sha3}, %
  keywordstyle=[1]\color{blue}\bfseries,
  keywords=[2]{address, bool, byte, bytes, bytes1, bytes2, bytes3, bytes4, bytes5, bytes6, bytes7, bytes8, bytes9, bytes10, bytes11, bytes12, bytes13, bytes14, bytes15, bytes16, bytes17, bytes18, bytes19, bytes20, bytes21, bytes22, bytes23, bytes24, bytes25, bytes26, bytes27, bytes28, bytes29, bytes30, bytes31, bytes32, enum, int, int8, int16, int24, int32, int40, int48, int56, int64, int72, int80, int88, int96, int104, int112, int120, int128, int136, int144, int152, int160, int168, int176, int184, int192, int200, int208, int216, int224, int232, int240, int248, int256, mapping, string, uint, uint8, uint16, uint24, uint32, uint40, uint48, uint56, uint64, uint72, uint80, uint88, uint96, uint104, uint112, uint120, uint128, uint136, uint144, uint152, uint160, uint168, uint176, uint184, uint192, uint200, uint208, uint216, uint224, uint232, uint240, uint248, uint256, var, void, ether, finney, szabo, wei, days, hours, minutes, seconds, weeks, years},  %
  keywordstyle=[2]\color{teal}\bfseries,
  keywords=[3]{block, blockhash, coinbase, difficulty, gaslimit, number, timestamp, msg, data, gas, sender, sig, now, tx, gasprice, origin},  %
  keywordstyle=[3]\color{violet}\bfseries,
  identifierstyle=\color{black},
  sensitive=false,
  comment=[l]{//},
  morecomment=[s]{/*}{*/},
  commentstyle=\color{gray}\ttfamily,
  stringstyle=\color{red}\ttfamily,
  morestring=[b]',
  morestring=[b]"
}
\newcommand{\KECCAK}{$\mathtt{KECCAK256}$\xspace}
\newcommand{\ECRECOVER}{$\mathtt{ECRECOVER}$\xspace}
\newcommand{\SHATWO}{$\mathtt{SHA256}$\xspace}
\newcommand{\RIPEMD}{$\mathtt{RIPEMD160}$\xspace}
\newcommand{\MODEXP}{$\mathtt{MODEXP}$\xspace}
\newcommand{\ECADD}{$\mathtt{ECADD}$\xspace}
\newcommand{\ECMUL}{$\mathtt{ECMUL}$\xspace}
\newcommand{\ECPAIRING}{$\mathtt{ECPAIRING}$\xspace}
\newcommand{\BLAKE}{$\mathtt{BLAKE2F}$\xspace}

\newcommand{\SSRFULL}{{\textit{Single-Contract Signature Replay}}\xspace}

\newcommand{\SFFULL}{{\textit{Signature Front-Running}}\xspace}
\newcommand{\SMFULL}{{\textit{Signature Malleability}}\xspace}
\newcommand{\ISVFULL}{{\textit{Insufficient Signature Verification}}\xspace}

\newcommand{\Precision}{95.4\%\xspace}
\newcommand{\Recall}{91.2\%\xspace}

\usepackage{enumerate}
\usepackage{graphics}
\usepackage{xparse} %
\usepackage{xspace}
\usepackage{multirow}

\usepackage{nicematrix}

\usepackage{booktabs}
\usepackage{multirow}
\usepackage{makecell}
\usepackage{enumitem}

\newcommand{\myparagraph}[1]{\textbf{#1.}\quad}

\newcommand{\numberedparagraph}[2]{\noindent \textbf{(#1) #2.} }

\newcommand{\etal}{{\textit{et al.}}\xspace}
\newcommand{\eg}{{\textit{e.g.}}\xspace}
\newcommand{\ie}{{\textit{i.e.}}\xspace}

\begin{document}

\title{Demystifying and Detecting Cryptographic Defects in Ethereum Smart Contracts}

\author{
	\IEEEauthorblockN{
		Jiashuo Zhang\IEEEauthorrefmark{1},
		Yiming Shen\IEEEauthorrefmark{2},
		Jiachi Chen\IEEEauthorrefmark{2}\IEEEauthorrefmark{5},
            Jianzhong Su\IEEEauthorrefmark{2}, 
		Yanlin Wang\IEEEauthorrefmark{2},\\
		Ting Chen\IEEEauthorrefmark{3},
  		Jianbo Gao\IEEEauthorrefmark{4}\IEEEauthorrefmark{5},
            Zhong Chen\IEEEauthorrefmark{1}\IEEEauthorrefmark{5}
	}
	\IEEEauthorblockA{\IEEEauthorrefmark{1}School of Computer Science, Peking University, Beijing, China}
 	\IEEEauthorblockA{\IEEEauthorrefmark{2}Sun Yat-sen University, Zhuhai, China} 
    \IEEEauthorblockA{\IEEEauthorrefmark{3}University of Electronic Science and Technology of China, Chengdu, China}
	\IEEEauthorblockA{\IEEEauthorrefmark{4}Beijing Key Laboratory of Security and Privacy in Intelligent Transportation,\\ Beijing Jiaotong University, Beijing, China} 
    \IEEEauthorblockA{\IEEEauthorrefmark{5}Corresponding Authors} 
	    \IEEEauthorblockA{
	zhangjiashuo@pku.edu.cn, seuilping@gmail.com, chenjch86@mail.sysu.edu.cn, sujzh3@mail2.sysu.edu.cn\\ yanlin-wang@outlook.com, brokendragon@uestc.edu.cn, gao@bjtu.edu.cn, zhongchen@pku.edu.cn}
 }

\maketitle
\begin{abstract}
Ethereum has officially provided a set of system-level cryptographic APIs to enhance smart contracts with cryptographic capabilities.
These APIs have been utilized in over 10\% of Ethereum transactions, motivating developers to implement various on-chain cryptographic tasks, such as digital signatures.
However, since developers may not always be cryptographic experts, their ad-hoc and potentially defective implementations could compromise the theoretical guarantees of cryptography, leading to real-world security issues.
To mitigate this threat, we conducted the first study aimed at demystifying and detecting cryptographic defects in smart contracts.
Through the analysis of 2,406 real-world security reports, we defined nine types of cryptographic defects in smart contracts with detailed descriptions and practical detection patterns. 
Based on this categorization, we proposed \tool, a fuzzing-based tool to automate the detection of cryptographic defects in smart contracts.
It combines transaction replaying and dynamic taint analysis to extract fine-grained crypto-related semantics and employs crypto-specific strategies to guide the test case generation process.
Furthermore, we collected a large-scale dataset containing 25,745 real-world crypto-related smart contracts and evaluated \tool's effectiveness on it.
The result demonstrated that \tool achieves an overall precision of 95.4\% and a recall of 91.2\%.
Notably, \tool revealed that 5,847 (22.7\%) out of 25,745 smart contracts contain at least one cryptographic defect, highlighting the prevalence of these defects.

\end{abstract}


\section{Introduction}

\label{sec:intro}

Cryptographic techniques, with their strong capabilities in securing data and communication, have demonstrated significant potential in enhancing the functionality of smart contracts~\cite{cryptocontracts,eip-712,belles2022circom}.
To prompt on-chain cryptographic practice, Ethereum has officially introduced a set of system-level cryptographic APIs~\cite{wood2014ethereum}, such as \ECRECOVER, to enable basic crypto operations within smart contracts.
These APIs effectively reduced the gas cost associated with complex cryptographic operations and prompted diverse on-chain cryptographic tasks such as \textit{digital signature}~\cite{eip-712} and \textit{Merkle proof}~\cite{merkle-proof}.
Currently, more than 10\% of Ethereum transactions use these crypto APIs~\cite{cryptocontracts}, highlighting the significance and prevalence of cryptographic practices in Ethereum smart contracts.

However, since smart contract developers may not necessarily be cryptographic experts, their implementation of cryptographic tasks could be error-prone.
Such defective implementations can compromise the theoretical security guarantees of cryptography and lead to real-world security issues in practice~\cite{defcon, nomadbridge, polygon}.
For example, a security team reported 52 smart contracts that suffered signature replay attacks~\cite{defcon}, illustrating the prevalence and damage of on-chain cryptographic defects.

Unfortunately, the community still lacks knowledge and tools to mitigate this threat.
A recent empirical study~\cite{cryptocontracts} revealed that 56.3\% of smart contract developers face obstacles in securing their cryptographic implementations, and 68.1\% of developers believe existing security tools need improvement to support their cryptographic practices.
Although many studies have focused on defects in smart contracts~\cite{chen2020defining, liu2018reguard, torres2018osiris, liu2022finding}, they mainly focus on issues arising from general programming tasks, such as \textit{Reentrancy}~\cite{liu2018reguard} and \textit{Integer Overflow}~\cite{torres2018osiris}, while rarely addressing defects specific to cryptographic practices.
Consequently, the characterization and mitigation of cryptographic defects remain an open challenge.

To bridge the gap, we conducted the first study focusing on demystifying and detecting cryptographic defects in Ethereum smart contracts.
To propose the definition and categorization of common cryptographic defects, we conducted an empirical study on 2,406 smart contract security reports from real-world security teams and investigated crypto-specific security issues they reported.
Based on an open-card sorting approach~\cite{wood2008card}, we introduced the first systematic taxonomy of cryptographic defects in smart contracts.
It includes nine categories of defects, covering common on-chain cryptographic tasks~\cite{cryptocontracts}, including \textit{digital signature}~\cite{nist1992digital}, \textit{Merkle proof}~\cite{merkle-proof}, \textit{message digest}~\cite{preneel1994cryptographic}, and \textit{random number generation}~\cite{random-number}.

Based on our defect definitions, we proposed \tool, a fuzzing-based approach to detect cryptographic defects in real-world smart contracts.
To the best of our knowledge, it is the first security technique targeting crypto-specific defects in contracts.
It integrates offline analysis with on-chain historical data to address the challenges posed by complicated cryptographic operations.
Specifically, \tool employs transaction replay and dynamic taint analysis to initialize the fuzzing context and extract essential crypto-related semantic information, such as data dependencies of cryptographic operations.
\tool utilizes a set of crypto-specific strategies to effectively generate test cases and exploit defects. These strategies guide \tool's test case generation with fine-grained semantic information and prevent it from getting stuck on trivial test cases, \ie, transactions directly reverted by cryptographic checks.
\tool executes the test cases and detects defects based on a set of crypto-specific oracles.
To evaluate \tool's effectiveness, we collected a dataset containing 25,745 real-world crypto-related contracts and ran \tool on it.
The results indicated that \tool achieves an overall precision of \Precision and a recall of \Recall.
Moreover, they demonstrated the prevalence of cryptographic defects in real-world contracts, revealing that 5,847 (22.7\%) out of these 25,745 contracts contain at least one defect.

We summarize our main contributions as follows:

\begin{itemize}
\item We conducted the first study on cryptographic defects in smart contracts.
Through the analysis of 2,406 security reports, we defined and categorized nine types of cryptographic defects, which expands the existing categorization of smart contract defects~\cite{chen2020defining,swclist}. We presented these defects with detailed descriptions and practical detection patterns to guide future security solutions.

\item We proposed \tool, the first tool to detect cryptographic defects in smart contracts.
It extracts fine-grained cryptographic semantics from on-chain data and employs crypto-specific strategies to guide the fuzzing process.
By addressing the functional gap of existing security tools, it has the potential to secure the emerging on-chain cryptographic practice.

\item We collected a large-scale dataset containing 25,745 real-world crypto-related smart contracts and evaluated \tool on it.
\tool revealed that 5,847 (22.7\%) of these contracts contain at least one cryptographic defect, with an overall precision of \Precision and a recall of \Recall.

\item We published the source code of \tool, all analysis results, and datasets at \href{https://github.com/Jiashuo-Zhang/CrySol}{\color{NavyBlue}{https://github.com/Jiashuo-Zhang/CrySol}}, to provide support to further studies.

\end{itemize}

\section{Background}
\label{sec:background}

\subsection{Ethereum Virtual Machine (EVM)}
Ethereum Virtual Machine (EVM) is the execution environment for Ethereum smart contracts~\cite{wood2014ethereum}.
It manages the on-chain states of smart contracts and transforms these states by iteratively executing instructions known as opcodes~\cite{evmopcode}.
The opcodes include stack/memory/storage operations, arithmetic calculations, and other functionalities required by smart contracts. 
For example, the $\mathtt{SLOAD}$ opcode reads a value from the contract's storage to the stack, and the $\mathtt{SSTORE}$ opcode writes a stack element to the storage.

Beyond the opcodes, Ethereum introduced several precompiled contracts as \textit{low-level extensions} of EVM~\cite{wood2014ethereum}.
They are implemented as built-in system-level contracts, to optimize the computation cost of specific functionalities, such as crypto operations.
User-defined contracts can use the $\mathtt{STATICCALL}$/$\mathtt{
	CALL}$/$\mathtt{
	CALLCODE}$/$\mathtt{
	DELEGATECALL}$ opcode to call precompiled contracts and execute their functionalities.

\subsection{Cryptographic APIs in EVM}
To enable cryptographic operations in smart contracts, Ethereum introduced nine cryptographic APIs to EVM~\cite{cryptocontracts}.
These APIs include one opcode (\KECCAK) and eight precompiled contracts (\ECRECOVER, \SHATWO, \RIPEMD, \MODEXP, \ECADD, \ECMUL, \ECPAIRING, \BLAKE).
Specifically, \KECCAK, \SHATWO, \RIPEMD, and \BLAKE provide four hash functions in smart contracts~\cite{wood2014ethereum, eip-152}, \ie, KECCAK256~\cite{keccak}, SHA2-256~\cite{penard2008secure}, RIPEMD-160~\cite{dobbertin1996ripemd}, and BLAKE2b~\cite{aumasson2014blake2}.
We collectively refer to these four APIs as hash operations in the remainder of this paper.
\ECRECOVER~\cite{wood2014ethereum} facilitates the on-chain verification of ECDSA signatures on the \textit{secp256k1} elliptic curve~\cite{johnson2001elliptic}.
\MODEXP~\cite{eip-198} enables big integer modular exponentiation.
\ECADD, \ECMUL, and \ECPAIRING ~\cite{eip-196,eip-197} provide elliptic operations of the \textit{alt\_bn\_128} curve to enable the verification of paring-based zero-knowledge proofs such as Groth16~\cite{groth2016size}.
 
These APIs largely reduce the gas cost of cryptographic operations and have thus attracted widespread application.
In a recent empirical study~\cite{cryptocontracts}, Zhang \etal found that 13.8\% of Ethereum transactions have utilized these crypto APIs.
In particular, \KECCAK, \ECRECOVER, \SHATWO APIs are the top three commonly used APIs, used by 13.0\%, 4.96\%, 0.56\% of transactions, respectively.

\subsection{Cryptographic Tasks in Smart Contracts} 
\label{crypto-practices}

Utilizing these crypto APIs, developers have implemented a variety of cryptographic tasks in smart contracts. Zhang \etal~\cite{cryptocontracts} analyzed the source codes of crypto-related smart contracts and classified common cryptographic tasks in smart contracts, including \textit{digital signatures} (used in 39.4\% of crypto-related contracts), \textit{vector commitments} (24.2\%), \textit{message digests} (17.4\%), and \textit{random number generators} (14.8\%).
We briefly introduce these tasks as follows:

\begin{itemize}[leftmargin=*]
    \item \textit{digital signatures}. Signatures are widely used for on-chain identity authentication~\cite{eip-712,eip-2612}. By combining \ECRECOVER with hash operations, developers can implement signature verification logic for ECDSA signatures~\cite{openzepplinECDSA}.
    \item \textit{vector commitments}. Vector commitments are widely used to enforce on-chain whitelist and other access control policies. They are typically implemented as Merkle proofs~\cite{merkle-proof}.
    \item \textit{message digest}. Message digest~\cite{preneel1994cryptographic} refers to the direct use of hash operations. It is commonly used to compute collision-resistant indexes for dynamic-length contents. 
    \item \textit{random number generator}. Random number generator~\cite{random-number} refers to generating pseudo-random numbers based on crypto operations. It is commonly used for on-chain gaming and gambling.
\end{itemize}

\subsection{Defects in Smart Contracts}
A software defect is an error, flaw, failure, or fault in a computer program or system that causes it to produce an incorrect or unexpected result, or to behave in unintended ways~\cite{florac1992software}.
Several previous studies have documented defects in smart contracts from different aspects~\cite{chen2020defining,chen2020survey,swclist}.
For example, Chen~\etal~\cite{chen2020defining} defined 20 types of defects in smart contract by analyzing StackExchange posts and real-world contracts. These defects impact the security, availability, performance, maintainability, and reusability of smart contracts.
With the ongoing innovation in on-chain applications, such as the integration with cryptographic techniques, and the ever-evolving security issues, the understanding and definition of contract defects are also evolving and expanding~\cite{chen2023healthier,swclist}.

\subsection{Security Reports for Smart Contracts}
Due to the prevalence of attacks, integrating security evaluations into smart contract development is essential~\cite{wan2021smart}.
Many third-party security teams, such as ConsenSys~\cite{Consensys} and Trails of Bits~\cite{trailsofbits}, offer security analysis services for smart contract projects.
They inspect the codes of smart contracts, search for defects, and produce detailed reports for developers.
These security reports, with comprehensive descriptions of real-world defects, are ideal information sources for defining smart contract defects.

\begin{table*}[t]
\setlength{\abovecaptionskip}{0cm}
\caption{Definitions of cryptographic defects in Ethereum smart contracts}
\resizebox{\textwidth}{!}{
\setlength\tabcolsep{3pt}

\begin{tabular}{@{}lcl@{}}
\toprule
\textbf{Cryptographic Defect}    & \textbf{ID}                                   & \textbf{Definition} \\ \midrule
Single-Contract Signature Replay & SSR          & Do not prevent the same signature from being used multiple times. \\
Cross-Contract Signature Replay & CSR              & Do not distinguish signatures for this contract from those for other contracts. \\
Signature Front-Running   &  SF                   &  Allow signatures in pending transactions to be front-run and preemptively used.         \\
Signature Malleability     &  SM                 &  Lack protection against signature malleability.          \\
Insufficient Signature Verification & ISV          & Do not properly check the result of signature verification.        \\
Merkle Proof Replay        &  MR                 &  Do not prevent the same Merkle proof from being used multiple times.         \\
Merkle Proof Front-Running  & MF                  & Allow Merkle proofs in pending transactions to be front-run and preemptively used.      \\
Hash Collisions With Dynamic-Length Arguments & HC & Do not prevent collisions when hashing concatenated dynamic-length arguments.         \\
Weak Randomness from Hashing Chain Attributes & WR &  Use the hash of chain attributes as randomness.        \\ \bottomrule
\end{tabular}
}

\label{table:defects}
\end{table*}

\section{Cryptographic Defects in Smart Contracts}

\label{sec:defects}
In this section, we conducted an empirical study on real-world security reports to define and categorize common cryptographic defects in Ethereum smart contracts.

\subsection{Data Collection}

During this process, we collected security reports from a wide range of real-world security teams.
Specifically, Etherscan~\cite{etherscansecurity} provides a list of 75 security teams that specialize in smart contract security. By searching on their official websites and accounts on social media platforms like Medium~\cite{medium}, we identified that 31 out of these security teams have publicly available security reports.
We manually collected these security reports and obtained a dataset containing 2,406 reports.

\subsection{Data Pre-processing}
To filter out security reports that relate to cryptographic practices, we conducted both keyword-based filtering and manual filtering on the collected reports.
During the keyword-based filtering, we utilized terms associated with cryptographic tasks %
such as ``signature'' and crypto API names like ``ecrecover'', as keywords.
As a result, we filtered out 893 reports that contained at least one keyword in their content.
However, due to the multiple meanings of keywords such as ``hash'', keyword-based methods are prone to inaccurate identification.
For example, the term ``hash'' in several reports actually refers to the commit hash of the code being audited.
We manually checked these reports to remove those with irrelevant keywords. Finally, we collected 211 crypto-related reports, which are available in our online supplementary material~\cite{supplement}.

\subsection{Data Analysis}
\begin{figure}[h]

    \centering
    \includegraphics[width=\linewidth]{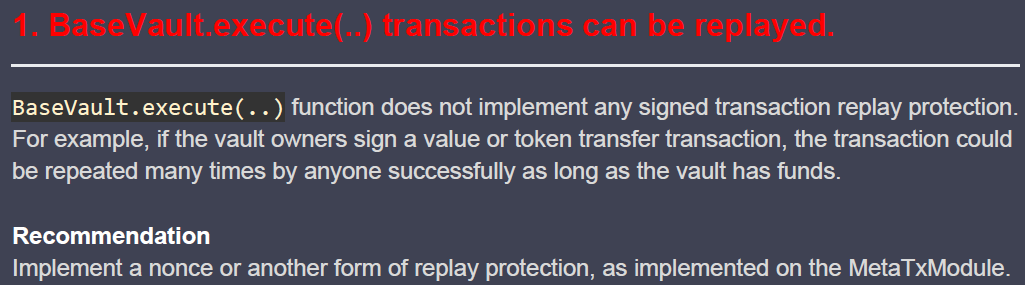}
    \vspace{-2em}
    \caption{An example of the cards of security reports.}
    \label{fig:report}

\end{figure}

We conducted a manual analysis on the collected 211 reports to investigate the categories of cryptographic defects in Ethereum smart contracts.
Due to the exploratory nature of our study, we did not introduce any pre-defined categories of defects.
Instead, we employed the open card sorting approach~\cite{wood2008card}, a common approach in software engineering for organizing information into logical groups, to define categories of defects.
In line with previous studies~\cite{cryptocontracts, chen2020defining, yang2023definition}, we created a card for each report, including detailed descriptions of the defects in the report and the root causes of them.
Fig.~\ref{fig:report} shows an example of the card for a security report~\cite{solidified-loopring}. 
It describes a defect in the \textit{BaseVault} contract that allows signature replay attacks. The root cause of this defect is the lack of protection against reused signatures, so that signatures in past transactions can be replayed multiple times.

During the card sorting procedure, two authors manually analyzed these cards to define the categories of cryptographic defects. For each card, they first examined its root cause to determine if it could be categorized under an existing category. If not, they evaluated the defect's representability and reproducibility to decide whether to introduce a new category. For example, defects that were highly specific to the business logic of a particular contract were not introduced as new categories.
After that, they engaged in discussions to resolve any disagreements and reached a consensus on the results.

\subsection{Defects Definition}
Through the analysis of security reports, we identified nine types of cryptographic defects in Ethereum smart contracts, covering all common on-chain cryptographic practices described in Section~\ref{crypto-practices}, \ie, \textit{digital signature}, \textit{vector commitment}, \textit{message digest} and \textit{random number generator}.
These defects could compromise theoretical security guarantees offered by cryptography and lead to unintended contract behaviors in practice.
Table~\ref{table:defects} enumerates each type of defect along with its definition. In the following, we describe these defects with detailed explanations and illustrative examples.

\begin{figure}[h]
\begin{lstlisting}
function permit(address owner, uint256 value, uint256 deadline, uint8 v,bytes32 r, bytes32 s) external {
    bytes32 hash = keccak256(abi.encode(owner, value, deadline));
    address signer = ecrecover(hash, v, r, s);
    require(signer != address(0),"Invalid Signature");
    require(owner == signer, "Invalid Signer");
    require(block.timestamp < deadline, "Permit Expired");
    _approve(owner, msg.sender, value);}
\end{lstlisting}
\vspace{-1em}
\caption{An example contract with SSR, CSR, and SF defects}
\vspace{-1em}

\label{example:sig}
\end{figure}

\numberedparagraph{1}{Single-Contract Signature Replay (SSR)}
Digital signatures are commonly used in smart contracts for on-chain access control~\cite{eip-2612, eip-712}. Transactions with valid signatures can perform sensitive operations in the contracts, such as transferring tokens. In such scenarios, a signature should be invalidated once it is verified, to prevent attackers from replaying the same signature and re-executing sensitive operations.
However, with this defect, the contract does not reject these valid but already used signatures.
Consequently, they may suffer from signature replay attacks: anyone who has observed valid signatures in past on-chain transactions can replay these signatures and pass the signature verification again.

\textbf{Example:} Fig.~\ref{example:sig} shows a defective
implementation of the ERC-20 permit function~\cite{eip-2612}. Ideally, this function should allow the \textit{msg.sender} to get approved to spend tokens after submitting a valid signature from the token owner (line 7).
However, since this function does not check whether each signature has been used, it always considers signatures used by past transactions as valid signatures. Consequently, it enables the replaying of signatures, allowing \textit{msg.sender} to gain repeated approvals for token spending.

\numberedparagraph{2}{Cross-Contract Signature Replay (CSR)}
This defect arises when two different contracts have an identical signing domain, \ie, the structures of their signed messages are exactly the same.
In such cases, a valid signature for contract \textit{A} will also be valid for contract \textit{B}, enabling cross-contract signature replay.
Compared to the SSR defect mentioned above, this defect involves a different attack vector:
SSR involves replaying historical signatures that previously used by the victim contract, while this defect involves replaying signatures from other contracts to the victim contract.
Both defects could lead to unauthorized access to sensitive operations.

\textbf{Example:} Take the permit function in Fig.~\ref{example:sig} as an example. Suppose there are two token contracts, \textit{A} and \textit{B}, each implementing the same permit function. The token owner, holding both tokens
\textit{A} and \textit{B}, intends to sign a permit for a spender of token \textit{A}. However, since the signed messages required by token \textit{A} and \textit{B} have exactly the same format (line 2-3), the signature intended solely for token \textit{A} also becomes valid for token \textit{B}. Consequently, a malicious spender can replay the same signature to token \textit{B} and successfully get approved, even though this was never intended.

\numberedparagraph{3}{Signature Front-Running (SF)}
In Ethereum, pending transactions, \ie, transactions that have been submitted to the network but not yet confirmed in a block, are publicly accessible~\cite{pendingtxns}. 
Therefore, signatures within pending transactions are susceptible to being captured and preemptively used in a front-running attack~\cite{daian2020flash,baum2022sok}.
This defect refers to situations where an attack transaction with front-run signatures can successfully pass the verification and lead to unintended contract behaviors.

\textbf{Example:} Consider the permit function in Fig.~\ref{example:sig}. Normally, a \textit{msg.sender} can obtain the approval by submitting a valid signature. However, in this case, an attacker can intercept the submitted signatures from pending transactions and initiate a front-running transaction to use them preemptively. If the attack succeeds, the new \textit{msg.sender} (\ie, the attacker), instead of the original \textit{msg.sender}, obtains the approval (line 7).

\numberedparagraph{4}{Signature Malleability (SM)}
The ECDSA signatures supported by \ECRECOVER precompiled contracts are malleable~\cite{groth2022security}. Specifically, given a valid signature $(v,r,s)$ for message $m$, anyone can generate another valid signature$(v’,r,s')$ for the same message $m$~\cite{decker2014bitcoin,swc-117}.
This defect refers to the lack of protection of signature malleability.
It is recognized to negatively impact the quality and maintainability of smart contracts~\cite{swc-117,openzepplinECDSA,thetanarena}, and can potentially lead to security issues such as signature replay attacks.

\textbf{Example:} Fig.~\ref{example:malleability} shows an example in which this defect can cause signature replay attacks.
In this case, the hash of the signature is used to prevent signature replay attacks (line 2-3).
Normally, after a signature is first verified, it is marked as \textit{used} (line 6), and any further attempts to use a \textit{used} signature are rejected (line 3).
However, due to the signature malleability, an attacker knowing a \textit{used} signature can generate a valid but unused signature for the same message.
Since the newly generated signature has not been marked as \textit{used} before, it can pass the check at line 3 and make a transfer again (line 7).

\begin{figure}[h]
\begin{lstlisting}
function transferWithSig(address to, uint256 value, uint8 v, bytes32 r, bytes32 s) public { 
    bytes32 sigHash = keccak256(v,r,s);
    require(!Used[sigHash]);
    address signer = ecrecover(keccak256(abi.encodePacked(to, value, address(this)),v,r,s));
    require(signer == owner);
    Used[sigHash] = true;
    transfer(to, value);
}
\end{lstlisting}
\vspace{-1em}
\caption{An example contract with the SM defect}
\label{example:malleability}
\vspace{-1em}
\end{figure}

\numberedparagraph{5}{Insufficient Signature Verification (ISV)}
Unlike standard signature verification process, which takes both the public key and the signature as input and indicates the signature's validity with a true/false output, \ECRECOVER employs the \textit{public key recovery} process~\cite{public-key-recovery} for signature verification, which only takes the signature as input and outputs the on-chain address of the ``expected'' signer.
As a result, when encountering an invalid signature, \ECRECOVER still returns an incorrect ``expected'' signer, instead of reverting the transaction.
Additionally, it simply returns zero if the signature is improperly formed~\cite{wood2014ethereum}. 
Therefore, when calling \ECRECOVER, contracts must check whether the returned ``expected'' signer is correct according to the business logic, \eg, by checking if it matches the token owner's address.
This defect arises when contracts do not properly verify \ECRECOVER's return value, leading to unintended contract behaviors.

 \textbf{Example:} Fig.~\ref{example:permit} illustrates an example of this defect. The intended behavior is to check the managers' signature before permitting an operation. However, attackers can submit a non-existent \textit{opType} and an improperly formed signature to make \ECRECOVER return zero. Since \textit{Manager[opType]} also defaults to zero for keys that don't exist, the attacker can successfully pass the signature verification (line 2-3) and gain unauthorized permission (line 4).

\begin{figure}[h]
\vspace{-1em}
\begin{lstlisting}
function permitOperation (address opType, uint256 opID, uint8 v, bytes32 r, bytes32 s) public { 
    address signer = ecrecover(keccak256(opType, opID), v, r, s);
    require(signer == Manager[opType]);
    permitted[opID]=true;
} 
\end{lstlisting}
\vspace{-1em}
\caption{An example contract with the ISV defect}
\label{example:permit}
\end{figure}

\numberedparagraph{6}{Merkle Proof Replay (MR)}
Merkle proofs are commonly employed to support on-chain whitelists and enable scenarios such as token airdropping~\cite{cryptocontracts}. Given a large set of users to be authorized, the contract owner can create a Merkle tree off-chain, distribute its leaves to the users, and upload the Merkle root in the contract~\cite{merkle-offchain}. Then, users can submit their leaves and the corresponding Merkle proofs to the contract. The contract will verify the Merkle proof before allowing users to do sensitive operations, such as minting NFTs.
Similar to signature replay attacks, lacking protection against Merkle proof replay can cause repeated/unauthorized access to sensitive operations.

\textbf{Example:} As illustrated in Fig.~\ref{example:merkle}, the contract allows whitelisted users to mint tokens (line 2) by submitting a valid Merkle proof (line 3). However, due to this defect, users, even those not in the whitelist, can replay past Merkle proofs submitted by whitelisted users and mint tokens.

\begin{figure}[h]
\begin{lstlisting}
function mint(string memory leaf, bytes32[] calldata merkleProof) external {
    if (MerkleProof.verify(merkleProof, merkleRoot, keccak256(abi.encodePacked(leaf)))){
     _mint(msg.sender, 1); }
}
\end{lstlisting}
\vspace{-1em}
\caption{An example contract with MR and MF defects.}
\label{example:merkle}
\vspace{-1em}
\end{figure}

\numberedparagraph{7}{Merkle Proof Front-Running (MF)}
This defect is similar to the \SFFULL defect. It allows attackers to capture Merkle proofs in the pending transactions and use them preemptively, which could enable unauthorized users to perform sensitive operations in the contract.

\textbf{Example:} The function in Fig.~\ref{example:merkle} also has this defect.
Specifically, anyone observing a pending Merkle proof can launch front-running attacks and preemptively mint tokens to their accounts.

\begin{figure}[h]
\vspace{-1em}
\begin{lstlisting}
 function addUsers(address[] calldata admins,address[] calldata regularUsers, bytes calldata signature) external {
    bytes32 hash = keccak256(abi.encodePacked(admins, regularUsers));
    address signer = hash.recover(signature);
    require(signer == owner);
    _addUser(admins,regularUsers)
}
\end{lstlisting}
\vspace{-1em}
\caption{An example contract with the HC defect.}

\label{example:hash}
\end{figure}

\numberedparagraph{8}{Hash Collisions With Dynamic-Length Arguments (HC)}
Crypto hash operations are expected to be collision-resistant~\cite{bellare1997collision}, \ie, it is computationally hard to find two input $a$ and $b$, \textit{s.t.}, $a\neq b \wedge hash(a)=hash(b)$. However, non-standard practice when hashing dynamic-length arguments, \ie, dynamic arrays in Solidity~\cite{abienodepacked}, could lead to ``collisions''.

\textbf{Example:} Fig.~\ref{example:hash} demonstrates this defect. 
Specifically, the built-in function \textit{abi.encodePacked} (line 2) packs all elements in order regardless of whether they're dynamic-length. Therefore, \KECCAK (\textit{abi.encodePacked} ([``0xa'', ``0xb''], [``0xc''])) is equal to \KECCAK(\textit{abi.encodePacked} ([``0xa''], [``0xb'', ``0xc''])), leading to a collision.
Consequently, attackers can rearrange the addresses in \textit{admins} and \textit{regularUsers} arrays, without changing the hash result (line 2). The signature verification still passes, but the content of these arrays and the contract's behavior (line 5) have been altered.

\numberedparagraph{9}{Weak Randomness from Hashing Chain Attributes (WR)}
Randomness is commonly used in scenarios such as on-chain gaming and gambling~\cite{cryptocontracts}. However, since there is a risk that miners could manipulate chain attributes such as \textit{block.timestamp} to their advantage~\cite{choi2021smartian}, generating random numbers by hashing chain attributes can compromise the security of these applications.

\textbf{Example:} Fig.~\ref{example:random} provides an example where this defect can be exploited to gain profits.
By choosing a \textit{block.timestamp} that meets the condition (line 4), a malicious miner can win the gambling game and receive the rewards (line 5).

\begin{figure}[h]
\begin{lstlisting}
function gamble() public payable {
    require(msg.value == 1 ether);
    uint8 rand = uint8(keccak256(block.timestamp, block.number))
    if (rand == 0) {
        msg.sender.transfer(2 ether);
}}
\end{lstlisting}
\vspace{-1em}
\caption{An example contract with the WR defect.}
\label{example:random}
\vspace{-0.5em}
\end{figure}

\noindent \myparagraph{Defect vs. Vulnerability vs. Bug}
We use the term \textit{defect} to collectively refer to the issues in cryptographic practices.
Compared to other terms such as \textit{vulnerability} and \textit{bug}, \textit{defect} has a wider scope~\cite{florac1992software,chen2020defining}, thus better representing these issues.
Specifically, \textit{vulnerability} refers to defects that can be directly exploited, while excluding other non-standard cryptographic implementations.
For example, while \SMFULL negatively impacts the quality and maintainability of the contract, it does not necessarily constitute a \textit{vulnerability}: it can only be directly exploited in certain cases like Fig.~\ref{example:malleability}.
Furthermore, \textit{bug} pertains to defects caused by coding errors.
However, defects like \SSRFULL are often a result of design flaws, \ie, the absence of a replay protection scheme, rather than coding errors.

\section{Methodology}

\label{sec:method}
Our results in Section~\ref{sec:defects} demonstrate nine defects of on-chain cryptographic practices.
To provide real-world evidence of these defects in Ethereum smart contracts and assist developers in detecting them in practice, we built \tool, an automated testing tool for Ethereum smart contracts.

\subsection{Design Decisions}
\tool is built on fuzzing, a plausible technique to detect defects in contracts~\cite{choi2021smartian,nguyen2020sfuzz}.
Compared to techniques like symbolic execution~\cite{luu2016making}, it can scale better to find defects with deep program paths and complex computations.
However, when applying fuzzing to crypto-related contracts, the inherent complexity of cryptography introduces new challenges.
In the following, we introduce these challenges and describe the design decisions we made to address these challenges.

\myparagraph{Properly Initializing the Fuzzing Context}
Crypto-related functions often involve intricate execution contexts. For example, to successfully call the function in Fig.~\ref{example:merkle}, the storage variable \textit{merkleRoot} needs to be properly initialized, and the transaction should include a valid Merkle proof pertaining to that specific \textit{merkleRoot}.
Common solutions, such as randomly initiating the contracts' states and transactions, may result in test transactions being trivially reverted by these cryptographic checks.
To overcome this, \tool utilizes real-world contracts' states and transactions to initialize the execution context of the fuzzing engine.
By integrating offline analysis with on-chain data, \tool provides a fuzzing context that is closer to real-world conditions, thereby improving the effectiveness and efficiency of the fuzzing process.

\myparagraph{Effectively Generating Test Cases}
Generating test cases that can exploit cryptographic defects requires certain guidance. For example, to exploit the SSR defect in Fig.~\ref{example:sig}, we need to construct two different transactions with the same signature. However, random methods could be highly inefficient to generate such test cases, since they require identifying which transaction parameters are included in the signed message.
To address this, \tool replays historical transactions of the contract and conducts dynamic taint analysis to extract crypto-related semantics, such as data-dependencies of cryptographic operations.
Utilizing these semantics, \tool employs a suite of crypto-specific strategies to effectively generate test cases that trigger potential defects.

\subsection{Overview}

\begin{figure}[t]
    \centering
\includegraphics[width=\columnwidth]{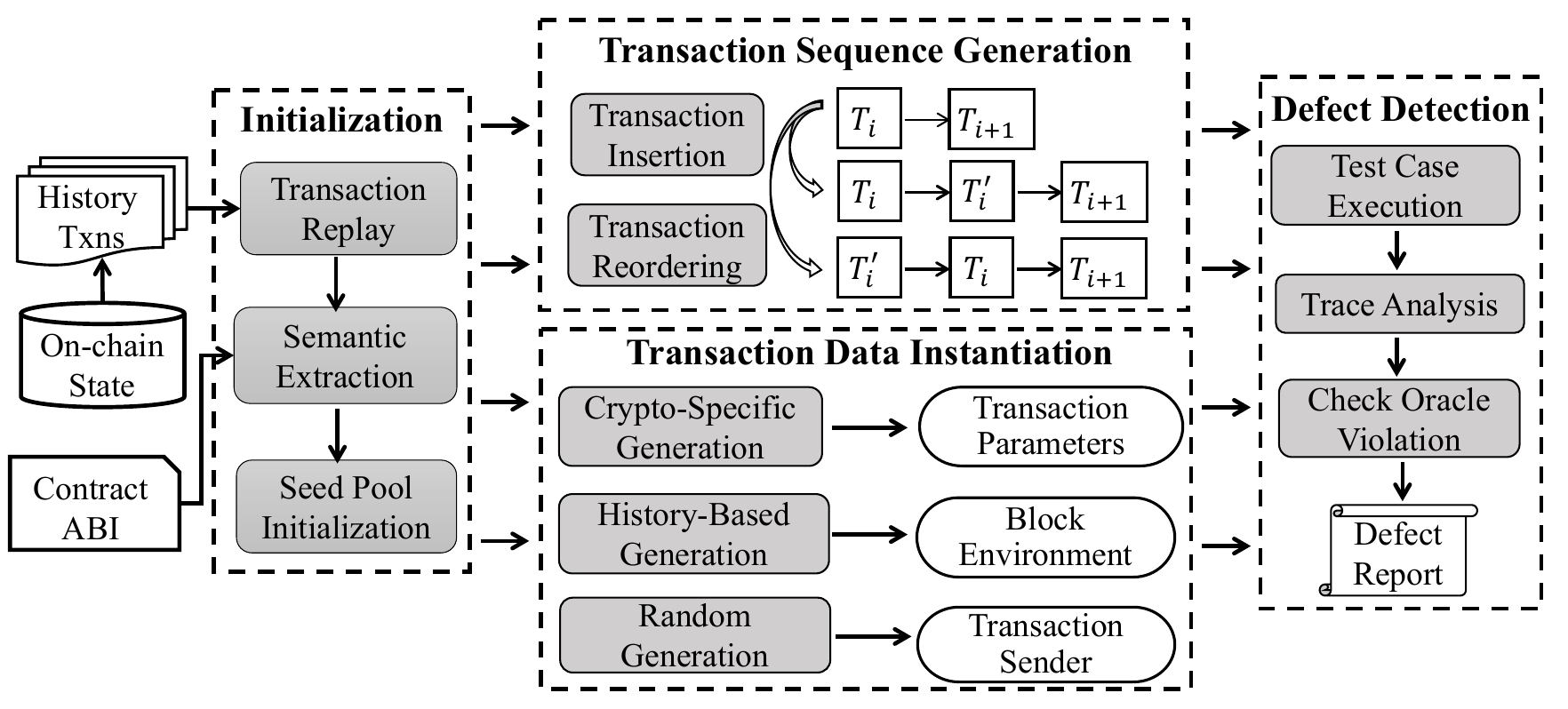}
    \vspace{-2em}
    \caption{The workflow of \tool.}
    \label{fig:overview}
    \vspace{-1em}
\end{figure}

Fig.~\ref{fig:overview} outlines the overall workflow of \tool.
Given a contract to analyze, \tool first replays its historical transactions to extract crypto-related semantics and initialize the seed pool (Section~\ref{sec:initialization}).
Then, \tool starts to generate test cases for the contract to trigger potential defects (Section~\ref{sec:sequence}). Specifically, the test case generation process involves two steps, \ie, generating the transaction sequence, and instantiating each transaction in the sequence with concrete parameter values.
Finally, \tool executes the test cases and analyzes the execution traces for defect detection (Section~\ref{sec:defectdetection}).

\subsection{Initialization}

\label{sec:initialization}

During the initialization, \tool replays historical on-chain transactions of the contract to collect crypto-related semantic information and initialize the seed pool for fuzzing.

\myparagraph{Transaction Replay}
\tool operates an Ethereum archive node~\cite{archivenode}, which retains all historical state information since the genesis block. For each transaction, \tool leverages an off-the-chain execution tool~\cite{kim2021off} designed for transaction replay to extract the contract's pre-state, \ie, the contract state before the transaction execution.
Then, \tool executes the transaction on this pre-state using an instrumented EVM, and gathers execution traces for subsequent analysis.

\myparagraph{Semantic Extraction}
Based on the execution traces, \tool collects the following crypto-related semantic information to guide the test case generation processes.
\begin{itemize}[leftmargin=*]
    \item \textit{Crypto-related functions}. To identify potential execution paths to trigger cryptographic defects, \tool analyzes the execution traces and filters out functions that used crypto operations. Specifically, for crypto APIs provided as precompiled contracts, \tool examines the destination address of all contract call opcodes ($\mathtt{STATICCALL}$, $\mathtt{
	CALL}$, $\mathtt{
	CALLCODE}$, and $\mathtt{
	DELEGATECALL}$) to determine whether the transaction calls these crypto APIs. For crypto APIs provided as opcode, \ie, \KECCAK, \tool analyzes all executed opcodes and checks whether there are crypto calls to \KECCAK.
    After identifying a function that uses cryptographic operations, \tool records all transactions traces of it for the subsequent data dependency analysis.
    
    \item \textit{Crypto-related data dependencies}. \tool employs dynamic taint analysis to extract data dependencies of the cryptographic operations. For example, to determine which transaction parameters may be an ECDSA signature, \tool marks slots of the transaction input data as sources and the arguments of \ECRECOVER as sinks. Then, it simulates taint propagation throughout the transaction's execution, checking if the sinks can be reached from the sources. For slots that can reach \ECRECOVER, it identifies to which parameters they correspond based on the contract's ABI and then marks these parameters as signature-related.
    Such information is essential for \tool to generate valid transactions that pass the cryptographic verification.
\end{itemize}

\myparagraph{Seed Pool Initialization}
After that, \tool initializes the fuzzing seed pool based on historical on-chain data.
Specifically, it includes all historical transactions of the contract as initial seeds.
Each seed contains all information required for executing the transaction: (1) the parameters and sender of the transaction;
(2) the pre-states of related contracts, including the contract directly called by the external transaction and other contracts called by internal transactions; and (3) the block environment, such as the block number and timestamp.

\subsection{Test Case Generation}
With the initialized seed pool and extracted semantic information, \tool begins to iteratively generate test cases.
Initially, it selects a seed from the seed pool and sets the contracts' pre-states and block environment recorded in the seed as the starting state for executing the generated test cases.
Based on the seed, \tool generates the test transaction sequence and instantiates each transaction with concrete input data.

\myparagraph{Transaction Sequence Generation}
\label{sec:sequence}
\tool supports two strategies to generate the transaction sequence, \ie, transaction insertion and transaction reordering.
When a seed is chosen, \tool includes the historical transaction from the seed into the initial transaction sequence.
Then, by strategically inserting new transactions to the initial sequence and re-ordering them, \tool generates a set of transaction sequences designed to exploit the defects.
For example, to exploit the SSR defect, \tool inserts a new attack transaction after the original historical transaction, calling the same function with a replayed signature.
To exploit the SF defect, \tool reorders the attack transaction to appear before the original transaction, enabling the front-running use of the signature.

\myparagraph{Transaction Data Instantiation}
This process is initiated when \tool needs to insert a new transaction into the transaction sequence.
Specifically, to instantiate the new transaction, \tool needs to generate three types of concrete data, \ie, the transaction parameters, transaction sender, and the block environment.
\tool generates these data based on the following three complementary strategies.
\label{sec:datamutation}

\begin{itemize}[leftmargin=*]
\item \textit{Crypto-Specific Generation.}
\tool employs a set of crypto-specific strategies to generate crypto-related transaction parameters.
Specifically, 
\tool analyzes crypto-related data dependencies and extracts crypto-related parameters that are used as the input of cryptographic operations.
Based on the analysis result, it strategically instantiates these parameters to exploit cryptographic defects.
For example, to exploit the SSR or SF defect, \tool needs to construct a new attack transaction containing the same signature as the original seed transaction.
To do so, it instantiates signature-related parameters by preserving their values in the original transactions, \ie, simulating the signature replay, while using history-based and random strategies to instantiate non-crypto-related parameters.

\item \textit{History-Based Generation.}
Given the security implications of cryptographic operations, crypto-related functions might operate within a more subtle context.
For example, a randomly selected transaction sender might fail to call the crypto-related functions due to the specific permission structure the contract initialized.
To better approximate real-world contexts, \tool offers the ability to instantiate transaction parameters or the transaction sender using values from all historical transactions calling the same function.
\item \textit{Random Generation.} In line with previous work~\cite{choi2021smartian,nguyen2020sfuzz,shou2023ityfuzz}, \tool infers parameter types based on the contract ABI specification~\cite{abispecification} and supports random generation of transaction parameters. Beyond transaction input data, \tool also supports randomly generating the transaction sender and the block environment.
\end{itemize}

By determining the sequence of transactions and instantiating each transaction with concrete parameter values, sender, and block environment, \tool generates test cases that can be concretely executed to exploit potential defects.

\subsection{Defects Detection}
\label{sec:defectdetection}
In the last phase, \tool executes the generated test cases and analyzes the execution traces for defect detection.
For each test case, \tool instantiates an instrumented EVM with the starting states of the test case, and executes the transaction sequence on it.
If the execution violates a pre-defined oracle, \tool reports the identified defect
along with the function containing the defect.
If the transaction involves multiple contracts, \tool also specifies the contract where the defect occurs by analyzing inter-contract calls during the transaction execution.
In the following, we describe the detailed oracles used by \tool to detect each type of defects.

\numberedparagraph{1}{Single-Contract Signature Replay (SSR)}
\tool examines the transaction sequence and checks whether it contains a successful signature replay attack.
Specifically, when a transaction calls \ECRECOVER, \tool searches for any subsequent transaction in the sequence that calls \ECRECOVER using the same parameters, \ie, replaying the signature. If the transaction with the replayed signature successfully executes and makes changes to the contract storage, \tool reports a SSR defect.
Additionally, we found that several token contracts intended to allow token minters to replay signatures and mint tokens until they reach the amount limit per address.
To reduce such false positives, \tool identifies such protective patterns by analyzing transaction execution traces.

\numberedparagraph{2}{Cross-Contract Signature Replay (CSR)}
\tool records calls to \ECRECOVER during the test case execution and checks whether each signed message includes the address of the contract that verifies the signature.
Specifically, \tool taints the return value of opcode $\mathtt{ADDRESS}$, which retrieves the contract's address. Then, \tool monitors if the taint flows into the $hash$ used in \ECRECOVER$(hash,v,r,s)$.
If the signed message does not include the contract's address, \ie, signatures for this contract are not distinguished from those of other contracts, \tool reports a CSR defect.

\numberedparagraph{3}{Signature Front-Running (SF)}
\tool examines the transaction sequence and checks whether it contains signature front-running attacks.
Specifically, \tool identifies cases where, given an original transaction that calls \ECRECOVER, there exists a preceding attack transaction from a different sender that calls \ECRECOVER with the same parameters.
If so, \tool conducts a differential analysis on the execution results of the original transaction and attack transaction.
It executes them based on the same start state respectively and compares the post-states after execution.
If the post-states differ, \ie, the attacker can make unintended changes to the contracts' states, \tool reports a SF defect.

\numberedparagraph{4}{Signature Malleability (SM)}
\tool analyzes whether there is protection against signature malleability.
Specifically, when encountering a call to \ECRECOVER$(hash,v,r,s)$, \tool analyzes the execution trace and checks whether a branching opcode ($\mathtt{JUMPI}$) is executed, conditioning on the comparison between $s$ and the constant elliptic curve order $secp256k1$~\cite{wood2014ethereum}.
If not, \ie, there is no protection against the signature malleability, \tool reports a SM defect.

\numberedparagraph{5}{Insufficient Signature Verification (ISV)}
\tool checks if there is a transaction that calls \ECRECOVER with parameters not used in any historical transactions, \ie, the signature is randomly forged by \tool.
If the transaction containing the forged signature successfully executes and makes changes to the storage, \tool reports an ISV defect.

\numberedparagraph{6}{Merkle Proof Replay (MR)}
\tool first identifies the verification process of Merkle proofs based on their operational characteristics. Specifically, \tool checks whether there is a sequence of hash operations during the transaction execution, where the input of the \textit{i}-th hash is the concatenation of the result of the \textit{(i-1)}-th hash and a proof element provided as the transaction parameters.
Then, similar to the detection of SSR, when encountering a transaction that verifies a Merkle proof, \tool searches for any subsequent transaction that replays that Merkle proof. If both transactions change the contract's storage, \tool reports a MR defect.
To reduce false positives, \tool identifies the same protective pattern for token minting as in the SSR defect.

\numberedparagraph{7}{Merkle Proof Front-Running (MF)}
The approach \tool uses to detect MF defects is analogous to the approach for SF defects.
Given an original transaction that verified the Merkle proofs, \tool searches for any preceding attack transaction that preemptively used the same Merkle proofs.
Then, a differential analysis is conducted on the execution results of these two transactions.
If the preceding transaction successfully executes and makes different changes to the contract storage, \tool reports a MF defect.

\numberedparagraph{8}{Hash Collisions With Dynamic-Length Arguments (HC)}
\tool conducts dynamic taint analysis on the input of each hash operation to detect HC defects. First, it determines which transaction parameters, if any, serve as input for these hash operations. Then, it checks whether these parameters are dynamic-length based on the contract's ABI.
If the hash input contains the concatenation of two dynamic-length parameters, \tool reports a HC defect.

\numberedparagraph{9}{Weak Randomness from Hashing Chain Attributes (WR)}
\tool leverages dynamic taint analysis to check whether the block attributes can affect hash operations. It first taints the returns of opcodes that acquire block attributes (\eg, $\mathtt{NUMBER}$ and $\mathtt{TIMESTAMP}$) and monitors whether the taints flow into hash operations.
If there is a hash operation that can be affected by chain attributes and the hash result determines a branch ($\mathtt{JUMPI}$) or storage operation ($\mathtt{SSTORE}$), \tool reports a WR defect.

\section{Evaluation}

\label{sec:evaluation}
The goal of the evaluation is two-fold. Firstly, we utilize a large-scale dataset containing 25,745 crypto-related smart contracts to evaluate the effectiveness of \tool in defect detection. Secondly,
by analyzing the results of this large-scale experiment,
we demystify cryptographic defects in the wild and gain insights into their prevalence and distribution.

\subsection{Evaluation Setup}

\myparagraph{Research Questions}Specifically, we focus on the following three research questions.

\begin{itemize}[leftmargin=*]
    \item \textbf{RQ1.} What is \tool's performance on our large-scale dataset? Can \tool find defects with high precision?
    \item \textbf{RQ2.} How effective of \tool in finding cryptographic defects in terms of recall? 
    \item \textbf{RQ3.} What is the prevalence and distribution of cryptographic defects in real-world smart contracts?
\end{itemize}

\myparagraph{Dataset}
To answer these research questions, we collected a large-scale dataset containing 25,745 real-world crypto-related smart contracts.
Specifically, using the same method as previous studies~\cite{cryptocontracts}, we first replayed 1,704,224,022 historical Ethereum transactions from block 1 to block 15,500,000 (from 2015.07 to 2022.09) and recorded the contracts that called crypto APIs.
In total, we identified 426,296 crypto-related contracts during the execution of historical transactions.
After that, we queried Etherscan~\cite{etherscan} to collect publicly available source codes and ABI information of these contracts.
As a result, we found 25,745 crypto-related smart contracts have available source codes and ABI information.
Among these contracts, 83.6\% of smart contracts have more than 10 historical transactions, suggesting that the majority of smart contracts in our dataset are engaged in real-world applications, rather than merely being toy contracts.

To retrieve contracts' historical transactions and states, \tool maintained an Ethereum archive node~\cite{archivenode} and recorded Ethereum on-chain raw states for subsequent analysis.
For each contract in the dataset, \tool fetched its historical transactions and states to initialize the fuzzing seed pool.
To ensure efficiency, the maximum seed pool size is set to be 500 transactions, which is considered adequate to cover common usage patterns of the contracts~\cite{liu2022finding,ye2023detecting}.
All experiments were conducted on a machine with two Intel Xeon(R) Platinum 8352V CPUs, 512 GB RAM, and running Ubuntu 22.04.2 LTS.

Our datasets, experiment outputs, and analysis results are all available in the supplementary materials~\cite{supplement}.

\subsection{RQ1: Detecting Defects in the Large-Scale Dataset}

To answer RQ1, we ran \tool on 25,745 smart contracts and analyzed the results. \tool took 408.0 hours to analyze 25,745 contracts, resulting in an average execution time of 57.1 seconds per contract. 
In total, \tool reported that 5,847 (22.7\%) contracts contain at least one defect.
Table~\ref{tab:defect} shows a breakdown of \tool's execution results for each defect type.

\begin{table}[t]
\setlength{\abovecaptionskip}{0cm}
    \begin{center}
        \caption{Cryptographic defects detected by \tool}
        \label{tab:defect}
        {
        \begin{tabular}{c||c|c|c|c|c}
            \hline
            \textbf{Defect}    & \textbf{\# Detected} &  \textbf{\# Sampled} & \textbf{\# TP} & \textbf{\# FP} & \textbf{Precision} \\
            \hline
            \textit{SSR}                  & 151              & 59 & 59 & 0 & 100.0\%         \\ %
            \hline
            \textit{CSR}            & 2,536              & 93 & 89 & 4 & 95.7\%            \\ %
            \hline
            \textit{SF}             & 274              & 71 & 69 & 2 & 97.2\%            \\ %
            \hline
            \textit{SM}           & 1,803               & 91 & 89 & 2 & 97.8\%        \\ %
            \hline
            \textit{ISV}           & 24              & 20 & 17 & 3 & 85.0\%           \\ %
            \hline
             \textit{MR}           & 122               & 54 & 48 & 6 & 88.9\%           \\ %
            \hline
             \textit{MF}           & 33                & 25 & 23 & 2 & 92.0\%           \\ %
            \hline
             \textit{HC}           & 89               & 46 & 43 & 3 & 93.5\%            \\ %
            \hline
             \textit{WR}           & 2,626                & 93 & 87 & 6 & 93.5\%            \\ %
            \hline
        \end{tabular}
        }
    \end{center}
   \vspace{-2em}
\end{table}

\myparagraph{Precision}
\label{sec:precision}
To evaluate the precision of \tool in detecting each type of defects, we manually analyzed the defects reported by \tool during the large-scale experiment.
In line with previous studies~\cite{yang2023definition,liu2021characterizing}, we randomly sampled a number of defects for each defect type to make the manual analysis feasible.
The sample size for each defect type was carefully chosen to achieve a confidence level of 95\% and a confidence interval of 10.
The second and third columns of Table~\ref{tab:defect} show the detected and sampled number of contracts with each defect, respectively.
Then, two of the authors independently labeled these contracts as true positives (TPs) or false positives (FPs), with the help of the third author to resolve any possible disagreements.
The fourth to sixth columns in Table~\ref{tab:defect} present the number of true positives, false positives, and the precision rate for each type of defect, respectively.
We then computed \tool's overall precision as a weighted average of these precision rates, with the weight being the number of each defect.
As a result, the overall precision of \tool is \Precision.

\myparagraph{False Positives}
After inspecting the false positives reported by \tool, we found that they are mainly caused by the following two factors.
The first is non-standard protective patterns in real-world contracts. For example, for \SMFULL, \tool reported a defect based on whether there is a condition that checks if the input \textit{s} for \ECRECOVER\textit{(hash,v,r,s)} is less than $secp256k1n/2$.
However, we found that some contracts used a non-standard protective pattern against signature malleability: they set the first bit of $s$ to 0 before using it as the actual input for \ECRECOVER\textit{(hash,v,r,s)}, thereby ensuring that $s$ is less than $secp256k1n/2$.
The second is the intended behavior of the contracts.
For example, for \ISVFULL, \tool checks whether a transaction with an invalid signature can successfully execute and make changes to the contract's storage.
However, we found that some smart contracts do not revert transactions when encountering invalid signatures. Instead, they intendedly record the signature verification results on-chain and continue to execute.
In such cases, the transaction with invalid signatures indeed results in the contract's storage changes, letting \tool falsely report an ISV defect.

\subsection{RQ2: Evaluating \tool on the Annotated Dataset}

To answer RQ2, we built an annotated dataset and evaluated the recall of \tool on it.
We have published the annotated dataset and analysis results in our online supplement materials~\cite{supplement}.

\myparagraph{Recall}
The evaluation of the recall requires a dataset with annotations of true positives and false negatives.
To establish the ground truth, we first randomly sampled a number of smart contracts from the large-scale dataset and manually annotated them.
Specifically, in line with previous studies~\cite{yang2023definition}, we randomly sampled 96 out of 25,745 contracts to achieve a confidence interval of 10 and a confidence level of 95\%.
Then, we followed the same labeling process as Section~\ref{sec:precision} to manually analyze these sampled contracts.
In total, we found 34 defects in these 96 contracts.
After comparing these manual labels and the results given by \tool, we found \tool reports 31 true positives, 1 false positive, and 3 false negatives for these contracts, which yields a recall of \Recall.

\myparagraph{False Negatives}
In detail, \tool failed to detect one SSR, one CSR, and one WR defect in 96 contracts.
After inspecting these false negatives, we found that they are mainly due to the lack of information to properly initialize the fuzzing context.
For example, while some smart contracts contain signature verification functionalities, such functions are rarely actually called. Consequently, \tool observed limited semantic information, hindering its ability to generate valid test cases for meaningful exploration.
However, automatically generating valid crypto-related transactions with solely off-line analysis is challenging.
In particular, cryptographic operations could render common techniques such as concolic testing~\cite{sen2007concolic} ineffective, since analyzing them results in complex symbolic expressions that cannot be handled by the SMT solver~\cite{corin2011efficient,vanhoef2018symbolic}.
Addressing these challenges is beyond the scope of this paper and is left as potential future work.

\subsection{RQ3: Characterizing Cryptographic Defects in the Wild}

While demonstrating the effectiveness of \tool, our large-scale experiment also provided a first close look at cryptographic defects in real-world smart contracts.

\begin{table}[t]
\setlength\tabcolsep{3pt}

\setlength{\abovecaptionskip}{0cm}
    \begin{center}
        \caption{The statistic metrics of defective contracts}
        \vspace{-1em}
        \label{tab:statistc}
        \resizebox{\columnwidth}{!}{
        \begin{NiceTabular}{c||c|c|c|c|c}
            \hline
            \textbf{Type} & \textbf{Prop.(\%)} & \textbf{LOC(avg)} & \textbf{\#Func(avg)} & \textbf{\#ETH(avg)} & \textbf{\#Txn(avg)} \\
            \hline
            \textit{SSR}  &  0.59\%             & 1369.5              &  33.8 & 6.0 & 8,163      \\ %
            \hline
            \textit{CSR}   &  9.85\%       & 1487.1              & 30.3  & 9.3 & 37,958          \\ %
            \hline
            \textit{SF}     & 1.06\%        & 1466.2             & 28.5 & 12.8 & 76,188       \\ %
            \hline
            \textit{SM}     & 7.00\%      & 1238.4             & 27.6  & 5.4 & 64,906       \\ %
            \hline
            \textit{ISV}      & 0.09\%     & 783.9              & 26.6 & 0.1 &  30,669           \\ %
            \hline
             \textit{MR}     & 0.47\%      & 1747.5              & 41.9  &  4.2 & 1,299          \\ %
            \hline
             \textit{MF}      & 0.13\%     &  1585.1          & 38.9  &  5.1  & 1,544      \\ %
            \hline
             \textit{HC}    & 0.35\%       & 1551.5              & 29.7  & 0.8 & 6,868          \\ %
            \hline
             \textit{WR}    & 10.20\%      & 1352.9              & 30.7  & 6.3 & 3,469         \\ %
              \hline\hline
            \textit{Total}    & 22.71\%      &  1396.8             & 30.3  & 7.1 &  22,930   \\ \hline
        \end{NiceTabular}
        }
    \end{center}
    \vspace{-2em}
\end{table}

\myparagraph{Prevalence and Distribution of Cryptographic Defects} 
The first column of Table~\ref{tab:statistc} presents the proportion of defective contracts regarding each defect type.
Among nine types of defects, WR, CSR, and SM are the most common, occurring in 2,626 (10.20\%), 2,536 (9.85\%), and 1,803 (7.00\%) of the analyzed smart contracts, respectively.
While the remaining six defect types are less common (appearing in about or less than 1\% of contracts), the total number of contracts affected by them is still considerable.
Such results provide real-world evidence for the findings of Zhang~\etal~\cite{cryptocontracts}, which suggest a lack of understanding of crypto-specific secure practices among smart contract developers. Note that a contract with cryptographic defects indicates deviations from best practices in cryptographic implementations. While defects may not directly lead to security issues, they can undermine the contract's maintainability and increase the risk of future security vulnerabilities. For instance, the CSR defect might not initially cause security problems when only one contract verifies the authorizer's signatures. However, if the system evolves and multiple contracts start using the same authorizer's signatures for managing sensitive operations, this defect can directly enable cross-contract signature replay attacks. A more detailed analysis of these cases is provided in our online supplementary materials~\cite{supplement}.

\myparagraph{Contracts with Cryptographic Defects}
To better understand cryptographic defects in the wild, we analyzed the average lines of code, number of external/public functions, Ether balances, and transaction counts of defective contracts, and presented them in columns three to six of Table~\ref{tab:statistc}.
The result shows that contracts with MR and MF defects are generally more complex than others, likely due to the inherent complexity of Merkle proofs and their applications, such as reward distribution.
Furthermore, contracts with SSR, CSR, SF, SM, and ISV defects, are more frequently called by real-world transactions, indicating a broader influence associated with signature-related defects.

\section{Discussion}

\subsection{Mitigations for Cryptographic Defects}

During the evaluation, we found that cryptographic defects are commonly caused by the direct use of low-level crypto APIs without necessary protection.
Therefore, in addition to introducing \tool, we provided possible solutions for each type of defect in Table~\ref{tab:solution}.
These solutions are summarized from the standard practices outlined in official Ethereum improvement proposals (EIPs)~\cite{eip-712,eip-2612} and defect remediation recommendations in security reports~\cite{solidified-loopring,pine,polygon,thetanarena}.

\begin{table}[t]
\setlength\tabcolsep{2pt}
\renewcommand{\arraystretch}{1.2}

\setlength{\abovecaptionskip}{0cm}
    \begin{center}
        \caption{Possible solutions for cryptographic defects}
        \vspace{-1em}
        \label{tab:solution}
        \resizebox{\columnwidth}{!}{
        \begin{tabular}{c||l}
            \hline
            \scriptsize \textbf{Type}    & \textbf{Possible Solution} \\
            \hline
            \textit{SSR}                  & Include a monotonic increasing nonce into the signed message\\
            \hline
            \textit{CSR}            & Include the contract address into the signed message       \\
            \hline
            \textit{SF}             & Prevent front-run signatures from causing unintended behaviors\\
            \hline
            \textit{SM}           & Add protection against ECDSA signature malleability\\
            \hline
            \textit{ISV}           & Check the return value of \ECRECOVER before sensitive operations                \\
            \hline
             \textit{MR}           & Check if the Merkle proof has been used before accepting it\\
            \hline
             \textit{MF}           &Prevent front-run Merkle proofs from causing unintended behaviors\\
            \hline
             \textit{HC}           &  Use collision-resistant encoding to hash dynamic-length variables          \\
            \hline
             \textit{WR}           & Use verifiable random function (VRF) for randomness            \\
            \hline
        \end{tabular}
        }
    \vspace{-2em}
    \end{center}
\end{table}

For example, Fig.~\ref{example:solution} shows a fixed version of the defective contract in Fig.~\ref{example:sig}.
It comes from a standard template~\cite{openzepplin-erc20permit} provided by OpenZeppelin~\cite{openzepplin}, which employs the above solutions to prevent SSR, CSR, and SF defects.
It integrates a nonce in the signed message to prevent SSR defects (line 4). It also includes a domain separator containing the contract address into the signed message to prevent the CSR defects (line 5).
Furthermore, to address SF defects, it replaces the address to be approved (line 9 in Fig.~\ref{example:solution} and line 7 in Fig.~\ref{example:sig}) from \textit{msg.sender} to the \textit{spender} specified by the signature (line 4). It ensures that even if an attacker front-runs the signature, he cannot change the intended contract behavior, \ie, \textit{owner} approving \textit{spender} for a certain \textit{value} of tokens.

In the supplementary material~\cite{supplement}, we provide more real-world examples to demonstrate how these solutions are applied to prevent cryptographic defects in practice.

\begin{figure}[h]
\begin{lstlisting}
function permit(address owner,address spender,uint256 value,uint256 deadline,uint8 v,bytes32 r,bytes32 s) public virtual {
    if (block.timestamp > deadline) {
        revert ERC2612ExpiredSignature(deadline);}
    bytes32 structHash = keccak256(abi.encode(PERMIT_TYPEHASH, owner, spender, value, _useNonce(owner), deadline));
    bytes32 hash = _hashTypedDataV4(structHash);
    address signer = ECDSA.recover(hash, v, r, s);
    if (signer != owner) {
        revert ERC2612InvalidSigner(signer, owner);}
    _approve(owner, spender, value);
}
\end{lstlisting}
\vspace{-1em}
\caption{Fixing defects in Fig.~\ref{example:sig}}
\label{example:solution}
\vspace{-1em}
\end{figure}

\subsection{Threats to Validity and Limitations}

\myparagraph{Threats to Validity}In the experiment, we employed random sampling to evaluate the effectiveness of \tool, which might introduce potential sampling bias.
To reduce the impact, we carefully selected the sampling ratio and size to achieve a confidence level of 95\% and a confidence interval of 10, which is considered sufficient in previous studies~\cite{liu2022finding, yang2023definition, chen2020defining, cryptocontracts}.
Additionally, we manually labeled true/false positives and negatives of the sampled contracts, which could potentially lead to labeling mistakes.
To mitigate this threat, we employed a double-check process, conducted by authors with more than three years of research experience in smart contract security.

\myparagraph{Limitations}
Despite \tool's strengths, it might have the following potential limitations.
First, \tool relies on pre-defined oracles to detect defects, which might not cover newly emerging defects beyond the existing nine categories of cryptographic defects.
However, given that these defined defects are derived from up to security reports from 31 security teams and involve all common on-chain cryptographic tasks, we believe \tool effectively captures \textit{common} cryptographic defects in existing smart contracts.
Its framework also allows future studies to easily incorporate new defects.
Second, \tool relies on on-chain information to guide the fuzzing process. In scenarios such as analyzing undeployed smart contracts, such information might not be directly accessible.
However, internal testing conducted before contract deployment, such as acceptance testing on local testnets, typically covers the main usage patterns of the contracts.
Utilizing these test transactions, \tool can extract necessary information and detect defects before deployment.

\section{Related Work}
\label{sec:rw}

\subsection{Defining and Detecting Defects in Smart Contracts}
Due to the recurring security incidents, a substantial body of research has been dedicated to defining and detecting defects in smart contracts~\cite{he2020smart, wang2021ethereum, ivanov2023security}.
Luu~\etal~\cite{luu2016making} took the first close look at smart contract security and proposed Oyente to detect four types of defects in smart contracts.
Chen~\etal~\cite{chen2020defining} defined 20 types of contract defects through the analysis of Stack Exchange posts and real-world smart contracts and proposed a tool to detect them~\cite{chen2021defectchecker}.
Liu~\etal~\cite{liu2022finding} studied access control bugs in smart contracts and detected them by dynamically role mining and conformance testing.
However, they mainly studied general programming defects such as \textit{Reentrancy}~\cite{liu2018reguard}, rather than crypto-specific defects we focused on.
For example, Liu~\etal~\cite{liu2022finding} focused on defective access control policies, rather than cryptographic defects that compromise the access control.
Ye~\etal~\cite{ye2023detecting} introduced a fuzzing tool to detect state inconsistency bugs, which utilizes contextual information collected from on-chain transactions to guide the fuzzing process.

While there is a lack of academic research on cryptographic defects, several defects we defined have attracted attention from the industry~\cite{swclist,defcon}.
To our knowledge, the Smart Contract Weakness Classification (SWC) list~\cite{swclist} has the most overlap with our categorization, which includes only four of nine defects we defined.
Specifically, 
SWC-121~\cite{swc-121} documents weaknesses caused by single-contract/cross-contract signature replays, involving SSR and CSR defects.
SWC-133~\cite{swc-133} and SWC-117~\cite{swc-117} are analogous to HC and SM defects, respectively.
While the SWC list documents these defects, it does not provide practical detect patterns or tools for their detection.

\subsection{Cryptographic Defects in Traditional Software}
Cryptographic defects have become a common cause of security issues in software~\cite{ami2022crypto,egele2013empirical,wickert2021python,zhang2022automatic}.
Lazar~\etal~\cite{lazar2014does} analyzed 269 crypto-related security incidents in the CVE database and found 83\% of them were caused by cryptographic defects introduced by developers' non-standard practices.
Egele~\etal~\cite{egele2013empirical} summarized six common cryptographic defects in Android applications and proposed a tool to detect them.
They found that 88\% of 11,748 Android applications that use cryptographic functionalities contain at least one defect.
Hazhirpasand~\etal~\cite{hazhirpasand2020java} found that 99.8\% of 489 Github projects using Java Cryptography Architecture (JCA) APIs contain at least one defect.

However, these studies mainly focus on cryptographic defects in traditional software.
Our results reveal differences between cryptographic defects in smart contracts and those in other well-studied software (\ie, Java applications), in terms of both definition and detection.
Firstly, due to the differences in common cryptographic tasks, the definition and categorization of defects in smart contracts differ inherently. For example, encryption-related defects are the most common in Java, but smart contracts rarely implement encryption, hence do not have these defects.
Secondly, the detection methods also differ.
In Java, defects often arise from direct API misuses, such as passing incorrect parameters to JCA APIs~\cite{hazhirpasand2020java}, and can be efficiently detected by static analyzers~\cite{zhang2022automatic}. However, detecting smart contract defects like SSR involves analyzing multiple transactions interacting with a stateful contract, making existing detection techniques difficult to apply.

\section{Conclusion}
\label{sec:conclusion}
In this paper, we conducted the first study aimed at understanding and uncovering cryptographic defects in smart contracts.
Through the analysis of 2,406 security reports, we proposed the first classification of cryptographic defects in smart contracts.
It encompasses nine distinct defect types and covers a wide range of cryptographic tasks in smart contracts.
To demonstrate these defects in real-world applications, we presented \tool, a fuzzing-based tool for cryptographic defect detection.
It collects fine-grained crypto-related semantics based on transaction replaying and dynamic taint analysis and incorporates crypto-specific fuzzing strategies for test case generation.
The evaluation results indicated that \tool can effectively detect real-world cryptographic defects, with an overall precision of \Precision and a recall of \Recall.
Furthermore, \tool revealed that 5,847 (22.7\%) out of 25,745 crypto-related smart contracts contain at least one cryptographic defect, demonstrating their prevalence in real-world cryptographic practices.


\bibliographystyle{IEEEtran}
\bibliography{ref}

\begin{thebibliography}{10}
\providecommand{\url}[1]{#1}
\csname url@samestyle\endcsname
\providecommand{\newblock}{\relax}
\providecommand{\bibinfo}[2]{#2}
\providecommand{\BIBentrySTDinterwordspacing}{\spaceskip=0pt\relax}
\providecommand{\BIBentryALTinterwordstretchfactor}{4}
\providecommand{\BIBentryALTinterwordspacing}{\spaceskip=\fontdimen2\font plus
\BIBentryALTinterwordstretchfactor\fontdimen3\font minus \fontdimen4\font\relax}
\providecommand{\BIBforeignlanguage}[2]{{%
\expandafter\ifx\csname l@#1\endcsname\relax
\typeout{** WARNING: IEEEtran.bst: No hyphenation pattern has been}%
\typeout{** loaded for the language `#1'. Using the pattern for}%
\typeout{** the default language instead.}%
\else
\language=\csname l@#1\endcsname
\fi
#2}}
\providecommand{\BIBdecl}{\relax}
\BIBdecl

\bibitem{cryptocontracts}
J.~Zhang, J.~Chen, Z.~Wan, T.~Chen, J.~Gao, and Z.~Chen, ``When contracts meets crypto: Exploring developers' struggles with ethereum cryptographic apis,'' in \emph{46th International Conference on Software Engineering (ICSE 24)}, 2024.

\bibitem{eip-712}
\BIBentryALTinterwordspacing
R.~Bloemen, L.~Logvinov, and J.~Evans, ``Eip-712: Typed structured data hashing and signing,'' 2017. [Online]. Available: \url{https://eips.ethereum.org/EIPS/eip-712}
\BIBentrySTDinterwordspacing

\bibitem{belles2022circom}
M.~Bell{\'e}s-Mu{\~n}oz, M.~Isabel, J.~L. Mu{\~n}oz-Tapia, A.~Rubio, and J.~Baylina, ``Circom: A circuit description language for building zero-knowledge applications,'' \emph{IEEE Transactions on Dependable and Secure Computing}, 2022.

\bibitem{wood2014ethereum}
G.~Wood \emph{et~al.}, ``Ethereum: A secure decentralised generalised transaction ledger,'' \emph{Ethereum project yellow paper}, vol. 151, no. 2014, pp. 1--32, 2014.

\bibitem{merkle-proof}
\BIBentryALTinterwordspacing
Ethereum, ``Merkle proofs for offline data integrity,'' 2023. [Online]. Available: \url{https://ethereum.org/vi/developers/tutorials/merkle-proofs-for-offline-data-integrity}
\BIBentrySTDinterwordspacing

\bibitem{defcon}
\BIBentryALTinterwordspacing
Z.~Bai, ``You may pay more than you can imagine,'' 2018. [Online]. Available: \url{https://github.com/nkbai/defcon26/tree/master/docs}
\BIBentrySTDinterwordspacing

\bibitem{nomadbridge}
\BIBentryALTinterwordspacing
Immunefi, ``Hack analysis: Nomad bridge, august 2022,'' 2022. [Online]. Available: \url{https://medium.com/immunefi/hack-analysis-nomad-bridge-august-2022-5aa63d53814a}
\BIBentrySTDinterwordspacing

\bibitem{polygon}
\BIBentryALTinterwordspacing
------, ``Polygon double-spend bugfix review,'' 2021. [Online]. Available: \url{https://medium.com/immunefi/polygon-double-spend-bug-fix-postmortem-2m-bounty-5a1db09db7f1}
\BIBentrySTDinterwordspacing

\bibitem{chen2020defining}
J.~Chen, X.~Xia, D.~Lo, J.~Grundy, X.~Luo, and T.~Chen, ``Defining smart contract defects on ethereum,'' \emph{IEEE Transactions on Software Engineering}, vol.~48, no.~1, pp. 327--345, 2020.

\bibitem{liu2018reguard}
C.~Liu, H.~Liu, Z.~Cao, Z.~Chen, B.~Chen, and B.~Roscoe, ``Reguard: finding reentrancy bugs in smart contracts,'' in \emph{Proceedings of the 40th International Conference on Software Engineering: Companion Proceeedings}, 2018, pp. 65--68.

\bibitem{torres2018osiris}
C.~F. Torres, J.~Sch{\"u}tte, and R.~State, ``Osiris: Hunting for integer bugs in ethereum smart contracts,'' in \emph{Proceedings of the 34th annual computer security applications conference}, 2018, pp. 664--676.

\bibitem{liu2022finding}
Y.~Liu, Y.~Li, S.-W. Lin, and C.~Artho, ``Finding permission bugs in smart contracts with role mining,'' in \emph{Proceedings of the 31st ACM SIGSOFT International Symposium on Software Testing and Analysis}, 2022, pp. 716--727.

\bibitem{wood2008card}
J.~R. Wood and L.~E. Wood, ``Card sorting: current practices and beyond,'' \emph{Journal of Usability Studies}, vol.~4, no.~1, pp. 1--6, 2008.

\bibitem{nist1992digital}
C.~Nist, ``The digital signature standard,'' \emph{Communications of the ACM}, vol.~35, no.~7, pp. 36--40, 1992.

\bibitem{preneel1994cryptographic}
B.~Preneel, ``Cryptographic hash functions,'' \emph{European Transactions on Telecommunications}, vol.~5, no.~4, pp. 431--448, 1994.

\bibitem{random-number}
\BIBentryALTinterwordspacing
Wikipedia, ``List of random number generators,'' 2023. [Online]. Available: \url{https://en.wikipedia.org/wiki/List_of_random_number_generators}
\BIBentrySTDinterwordspacing

\bibitem{swclist}
\BIBentryALTinterwordspacing
S.~Registry, ``Smart contract weakness classification and test cases,'' 2023. [Online]. Available: \url{https://swcregistry.io/}
\BIBentrySTDinterwordspacing

\bibitem{evmopcode}
\BIBentryALTinterwordspacing
Ethereum, ``Opcodes for the evm,'' 2023. [Online]. Available: \url{https://ethereum.org/en/developers/docs/evm/opcodes}
\BIBentrySTDinterwordspacing

\bibitem{eip-152}
\BIBentryALTinterwordspacing
H.~Tjaden, L.~Matt, D.~Piotr, and H.~James, ``Eip-152: Add blake2 compression function `f` precompile,'' 2016. [Online]. Available: \url{https://eips.ethereum.org/EIPS/eip-152}
\BIBentrySTDinterwordspacing

\bibitem{keccak}
\BIBentryALTinterwordspacing
B.~Guido, D.~Joan, P.~Michal, and G.~V. Assche, ``The keccak sha-3 submission,'' 2011. [Online]. Available: \url{https://keccak.team/files/ Keccak-submission-3.pdf}
\BIBentrySTDinterwordspacing

\bibitem{penard2008secure}
W.~Penard and T.~van Werkhoven, ``On the secure hash algorithm family,'' \emph{Cryptography in context}, pp. 1--18, 2008.

\bibitem{dobbertin1996ripemd}
H.~Dobbertin, A.~Bosselaers, and B.~Preneel, ``Ripemd-160: A strengthened version of ripemd,'' in \emph{Fast Software Encryption: Third International Workshop Cambridge, UK, February 21--23 1996 Proceedings 3}.\hskip 1em plus 0.5em minus 0.4em\relax Springer, 1996, pp. 71--82.

\bibitem{aumasson2014blake2}
J.-P. Aumasson, W.~Meier, R.~C.-W. Phan, L.~Henzen, J.-P. Aumasson, W.~Meier, R.~C.-W. Phan, and L.~Henzen, ``Blake2,'' \emph{The Hash Function BLAKE}, pp. 165--183, 2014.

\bibitem{johnson2001elliptic}
D.~Johnson, A.~Menezes, and S.~Vanstone, ``The elliptic curve digital signature algorithm (ecdsa),'' \emph{International Journal of Information Security}, vol.~1, pp. 36--63, 2001.

\bibitem{eip-198}
\BIBentryALTinterwordspacing
V.~Buterin, ``Eip-198: Big integer modular exponentiation,'' 2017. [Online]. Available: \url{https://eips.ethereum.org/EIPS/eip-198}
\BIBentrySTDinterwordspacing

\bibitem{eip-196}
\BIBentryALTinterwordspacing
C.~Reitwiessner, ``Eip-196: Precompiled contracts for addition and scalar multiplication on the elliptic curve alt\_bn128,'' 2017. [Online]. Available: \url{https://eips.ethereum.org/EIPS/eip-196}
\BIBentrySTDinterwordspacing

\bibitem{eip-197}
\BIBentryALTinterwordspacing
V.~Buterin and C.~Reitwiessner, ``Eip-197: Precompiled contracts for optimal ate pairing check on the elliptic curve alt\_bn128,'' 2017. [Online]. Available: \url{https://eips.ethereum.org/EIPS/eip-197}
\BIBentrySTDinterwordspacing

\bibitem{groth2016size}
J.~Groth, ``On the size of pairing-based non-interactive arguments,'' in \emph{Advances in Cryptology--EUROCRYPT 2016: 35th Annual International Conference on the Theory and Applications of Cryptographic Techniques, Vienna, Austria, May 8-12, 2016, Proceedings, Part II 35}.\hskip 1em plus 0.5em minus 0.4em\relax Springer, 2016, pp. 305--326.

\bibitem{eip-2612}
\BIBentryALTinterwordspacing
M.~Lundfal, ``Erc-2612: Permit extension for eip-20 signed approvals,'' 2020. [Online]. Available: \url{https://eips.ethereum.org/EIPS/eip-2612}
\BIBentrySTDinterwordspacing

\bibitem{openzepplinECDSA}
\BIBentryALTinterwordspacing
Openzepplin, ``Checking signatures on-chain,'' 2023. [Online]. Available: \url{https://docs.openzeppelin.com/contracts/2.x/utilities}
\BIBentrySTDinterwordspacing

\bibitem{florac1992software}
W.~A. Florac \emph{et~al.}, \emph{Software quality measurement: A framework for counting problems and defects}.\hskip 1em plus 0.5em minus 0.4em\relax Carnegie Mellon University, Software Engineering Institute, 1992.

\bibitem{chen2020survey}
H.~Chen, M.~Pendleton, L.~Njilla, and S.~Xu, ``A survey on ethereum systems security: Vulnerabilities, attacks, and defenses,'' \emph{ACM Computing Surveys (CSUR)}, vol.~53, no.~3, pp. 1--43, 2020.

\bibitem{chen2023healthier}
J.~Chen, M.~Huang, Z.~Lin, P.~Zheng, and Z.~Zheng, ``To healthier ethereum: A comprehensive and iterative smart contract weakness enumeration,'' \emph{arXiv preprint arXiv:2308.10227}, 2023.

\bibitem{wan2021smart}
Z.~Wan, X.~Xia, D.~Lo, J.~Chen, X.~Luo, and X.~Yang, ``Smart contract security: a practitioners' perspective,'' in \emph{2021 IEEE/ACM 43rd International Conference on Software Engineering (ICSE)}.\hskip 1em plus 0.5em minus 0.4em\relax IEEE, 2021, pp. 1410--1422.

\bibitem{Consensys}
\BIBentryALTinterwordspacing
Consensys, ``A complete suite of products to create and participate in web3,'' 2023. [Online]. Available: \url{https://consensys.io/}
\BIBentrySTDinterwordspacing

\bibitem{trailsofbits}
\BIBentryALTinterwordspacing
T.~of~Bits, ``Trails of bits,'' 2023. [Online]. Available: \url{https://www.trailofbits.com/}
\BIBentrySTDinterwordspacing

\bibitem{etherscansecurity}
\BIBentryALTinterwordspacing
Etherscan, ``Smart contracts audit and security,'' 2023. [Online]. Available: \url{https://etherscan.io/directory/Smart_Contracts/Smart_Contracts_Audit_And_Security}
\BIBentrySTDinterwordspacing

\bibitem{medium}
\BIBentryALTinterwordspacing
Medium, ``Medium,'' 2023. [Online]. Available: \url{https://medium.com/}
\BIBentrySTDinterwordspacing

\bibitem{supplement}
\BIBentryALTinterwordspacing
CrySol, ``Online supplement material,'' 2023. [Online]. Available: \url{https://github.com/Jiashuo-Zhang/CrySol}
\BIBentrySTDinterwordspacing

\bibitem{yang2023definition}
S.~Yang, J.~Chen, and Z.~Zheng, ``Definition and detection of defects in nft smart contracts,'' in \emph{32nd ACM SIGSOFT International Symposium on Software Testing and Analysis}, 2023.

\bibitem{solidified-loopring}
\BIBentryALTinterwordspacing
Solidified, ``Audit report for loopring on may 21st, 2020.'' 2020. [Online]. Available: \url{https://github.com/solidified-platform/audits/blob/master/Audit Report - Loopring Hebao Wallet [21.05.2020].pdf}
\BIBentrySTDinterwordspacing

\bibitem{pendingtxns}
\BIBentryALTinterwordspacing
Etherscan, ``Ethereum pending transactions,'' 2023. [Online]. Available: \url{https://etherscan.io/txsPending}
\BIBentrySTDinterwordspacing

\bibitem{daian2020flash}
P.~Daian, S.~Goldfeder, T.~Kell, Y.~Li, X.~Zhao, I.~Bentov, L.~Breidenbach, and A.~Juels, ``Flash boys 2.0: Frontrunning in decentralized exchanges, miner extractable value, and consensus instability,'' in \emph{2020 IEEE Symposium on Security and Privacy (SP)}.\hskip 1em plus 0.5em minus 0.4em\relax IEEE, 2020, pp. 910--927.

\bibitem{baum2022sok}
C.~Baum, J.~Hsin-yu Chiang, B.~David, T.~K. Frederiksen, and L.~Gentile, ``Sok: Mitigation of front-running in decentralized finance,'' in \emph{International Conference on Financial Cryptography and Data Security}.\hskip 1em plus 0.5em minus 0.4em\relax Springer, 2022, pp. 250--271.

\bibitem{groth2022security}
J.~Groth and V.~Shoup, ``On the security of ecdsa with additive key derivation and presignatures,'' in \emph{Annual International Conference on the Theory and Applications of Cryptographic Techniques}.\hskip 1em plus 0.5em minus 0.4em\relax Springer, 2022, pp. 365--396.

\bibitem{decker2014bitcoin}
C.~Decker and R.~Wattenhofer, ``Bitcoin transaction malleability and mtgox,'' in \emph{Computer Security-ESORICS 2014: 19th European Symposium on Research in Computer Security, Wroclaw, Poland, September 7-11, 2014. Proceedings, Part II 19}.\hskip 1em plus 0.5em minus 0.4em\relax Springer, 2014, pp. 313--326.

\bibitem{swc-117}
\BIBentryALTinterwordspacing
S.~Registry, ``Signature malleability,'' 2023. [Online]. Available: \url{https://swcregistry.io/docs/SWC-117}
\BIBentrySTDinterwordspacing

\bibitem{thetanarena}
\BIBentryALTinterwordspacing
Verichains, ``Verichains public audit report - thetanarena,'' 2021. [Online]. Available: \url{https://github.com/verichains/public-audit-reports/blob/main/Verichains Public Audit Report - ThetanArena - v1.2.pdf}
\BIBentrySTDinterwordspacing

\bibitem{public-key-recovery}
\BIBentryALTinterwordspacing
E.~C. D.~S. Algorithm, ``Public key recovery,'' 2023. [Online]. Available: \url{https://en.wikipedia.org/wiki/Elliptic_Curve_Digital_Signature_Algorithm#Public_key_recovery}
\BIBentrySTDinterwordspacing

\bibitem{merkle-offchain}
\BIBentryALTinterwordspacing
E.~O. Documentation, ``Merkle proofs for offline data integrity,'' 2023. [Online]. Available: \url{https://ethereum.org/es/developers/tutorials/merkle-proofs-for-offline-data-integrity}
\BIBentrySTDinterwordspacing

\bibitem{bellare1997collision}
M.~Bellare and P.~Rogaway, ``Collision-resistant hashing: Towards making uowhfs practical,'' in \emph{Annual International Cryptology Conference}.\hskip 1em plus 0.5em minus 0.4em\relax Springer, 1997, pp. 470--484.

\bibitem{abienodepacked}
\BIBentryALTinterwordspacing
Ethereuk, ``Non-standard packed mode,'' 2023. [Online]. Available: \url{https://docs.soliditylang.org/en/v0.8.23/abi-spec.html#non-standard-packed-mode}
\BIBentrySTDinterwordspacing

\bibitem{choi2021smartian}
J.~Choi, D.~Kim, S.~Kim, G.~Grieco, A.~Groce, and S.~K. Cha, ``Smartian: Enhancing smart contract fuzzing with static and dynamic data-flow analyses,'' in \emph{2021 36th IEEE/ACM International Conference on Automated Software Engineering (ASE)}.\hskip 1em plus 0.5em minus 0.4em\relax IEEE, 2021, pp. 227--239.

\bibitem{nguyen2020sfuzz}
T.~D. Nguyen, L.~H. Pham, J.~Sun, Y.~Lin, and Q.~T. Minh, ``sfuzz: An efficient adaptive fuzzer for solidity smart contracts,'' in \emph{Proceedings of the ACM/IEEE 42nd International Conference on Software Engineering}, 2020, pp. 778--788.

\bibitem{luu2016making}
L.~Luu, D.-H. Chu, H.~Olickel, P.~Saxena, and A.~Hobor, ``Making smart contracts smarter,'' in \emph{Proceedings of the 2016 ACM SIGSAC conference on computer and communications security}, 2016, pp. 254--269.

\bibitem{archivenode}
\BIBentryALTinterwordspacing
Ethereum, ``Ethereum archive node,'' 2023. [Online]. Available: \url{https://ethereum.org/en/developers/docs/nodes-and-clients/archive-nodes}
\BIBentrySTDinterwordspacing

\bibitem{kim2021off}
Y.~Kim, S.~Jeong, K.~Jezek, B.~Burgstaller, and B.~Scholz, ``An off-the-chain execution environment for scalable testing and profiling of smart contracts,'' in \emph{2021 USENIX Annual Technical Conference (USENIX ATC 21)}, 2021, pp. 565--579.

\bibitem{shou2023ityfuzz}
C.~Shou, S.~Tan, and K.~Sen, ``Ityfuzz: Snapshot-based fuzzer for smart contract,'' in \emph{Proceedings of the 32nd ACM SIGSOFT International Symposium on Software Testing and Analysis}, 2023, pp. 322--333.

\bibitem{abispecification}
\BIBentryALTinterwordspacing
S.~Documentation, ``Contract abi specification,'' 2023. [Online]. Available: \url{https://docs.soliditylang.org/en/latest/abi-spec.html}
\BIBentrySTDinterwordspacing

\bibitem{etherscan}
\BIBentryALTinterwordspacing
Etherscan, ``The ethereum blockchain explorer,'' 2023. [Online]. Available: \url{https://etherscan.io/}
\BIBentrySTDinterwordspacing

\bibitem{ye2023detecting}
M.~Ye, Y.~Nan, Z.~Zheng, D.~Wu, and H.~Li, ``Detecting state inconsistency bugs in dapps via on-chain transaction replay and fuzzing,'' in \emph{Proceedings of the 32nd ACM SIGSOFT International Symposium on Software Testing and Analysis}, 2023, pp. 298--309.

\bibitem{liu2021characterizing}
L.~Liu, L.~Wei, W.~Zhang, M.~Wen, Y.~Liu, and S.-C. Cheung, ``Characterizing transaction-reverting statements in ethereum smart contracts,'' in \emph{2021 36th IEEE/ACM International Conference on Automated Software Engineering (ASE)}.\hskip 1em plus 0.5em minus 0.4em\relax IEEE, 2021, pp. 630--641.

\bibitem{sen2007concolic}
K.~Sen, ``Concolic testing,'' in \emph{Proceedings of the 22nd IEEE/ACM international conference on Automated software engineering}, 2007, pp. 571--572.

\bibitem{corin2011efficient}
R.~Corin and F.~A. Manzano, ``Efficient symbolic execution for analysing cryptographic protocol implementations,'' in \emph{International Symposium on Engineering Secure Software and Systems}.\hskip 1em plus 0.5em minus 0.4em\relax Springer, 2011, pp. 58--72.

\bibitem{vanhoef2018symbolic}
M.~Vanhoef and F.~Piessens, ``Symbolic execution of security protocol implementations: Handling cryptographic primitives,'' in \emph{12th USENIX Workshop on Offensive Technologies (WOOT 18)}, 2018.

\bibitem{pine}
\BIBentryALTinterwordspacing
Quantstamp, ``Pine audit report,'' 2022. [Online]. Available: \url{https://certificate.quantstamp.com/full/pine.pdf}
\BIBentrySTDinterwordspacing

\bibitem{openzepplin-erc20permit}
\BIBentryALTinterwordspacing
Openzepplin, ``Implementation of the erc-20 permit extension,'' 2023. [Online]. Available: \url{https://github.com/OpenZeppelin/openzeppelin-contracts/blob/master/contracts/token/ERC20/extensions/ERC20Permit.sol}
\BIBentrySTDinterwordspacing

\bibitem{openzepplin}
\BIBentryALTinterwordspacing
------, ``The standard for secure blockchain applications,'' 2023. [Online]. Available: \url{https://www.openzeppelin.com/}
\BIBentrySTDinterwordspacing

\bibitem{he2020smart}
D.~He, Z.~Deng, Y.~Zhang, S.~Chan, Y.~Cheng, and N.~Guizani, ``Smart contract vulnerability analysis and security audit,'' \emph{IEEE Network}, vol.~34, no.~5, pp. 276--282, 2020.

\bibitem{wang2021ethereum}
Z.~Wang, H.~Jin, W.~Dai, K.-K.~R. Choo, and D.~Zou, ``Ethereum smart contract security research: survey and future research opportunities,'' \emph{Frontiers of Computer Science}, vol.~15, pp. 1--18, 2021.

\bibitem{ivanov2023security}
N.~Ivanov, C.~Li, Q.~Yan, Z.~Sun, Z.~Cao, and X.~Luo, ``Security threat mitigation for smart contracts: A comprehensive survey,'' \emph{ACM Computing Surveys}, 2023.

\bibitem{chen2021defectchecker}
J.~Chen, X.~Xia, D.~Lo, J.~Grundy, X.~Luo, and T.~Chen, ``Defectchecker: Automated smart contract defect detection by analyzing evm bytecode,'' \emph{IEEE Transactions on Software Engineering}, vol.~48, no.~7, pp. 2189--2207, 2021.

\bibitem{swc-121}
\BIBentryALTinterwordspacing
S.~Registry, ``Missing protection against signature replay attacks,'' 2023. [Online]. Available: \url{https://swcregistry.io/docs/SWC-121}
\BIBentrySTDinterwordspacing

\bibitem{swc-133}
\BIBentryALTinterwordspacing
SWC, ``Hash collisions with multiple variable length arguments,'' 2023. [Online]. Available: \url{https://swcregistry.io/docs/SWC-133}
\BIBentrySTDinterwordspacing

\bibitem{ami2022crypto}
A.~S. Ami, N.~Cooper, K.~Kafle, K.~Moran, D.~Poshyvanyk, and A.~Nadkarni, ``Why crypto-detectors fail: A systematic evaluation of cryptographic misuse detection techniques,'' in \emph{2022 IEEE Symposium on Security and Privacy (SP)}.\hskip 1em plus 0.5em minus 0.4em\relax IEEE, 2022, pp. 614--631.

\bibitem{egele2013empirical}
M.~Egele, D.~Brumley, Y.~Fratantonio, and C.~Kruegel, ``An empirical study of cryptographic misuse in android applications,'' in \emph{Proceedings of the 2013 ACM SIGSAC Conference on Computer \& Communications Security}, 2013, pp. 73--84.

\bibitem{wickert2021python}
A.-K. Wickert, L.~Baumg{\"a}rtner, F.~Breitfelder, and M.~Mezini, ``Python crypto misuses in the wild,'' in \emph{Proceedings of the 15th ACM/IEEE International Symposium on Empirical Software Engineering and Measurement (ESEM)}, 2021, pp. 1--6.

\bibitem{zhang2022automatic}
Y.~Zhang, M.~M.~A. Kabir, Y.~Xiao, D.~Yao, and N.~Meng, ``Automatic detection of java cryptographic api misuses: Are we there yet?'' \emph{IEEE Transactions on Software Engineering}, vol.~49, no.~1, pp. 288--303, 2022.

\bibitem{lazar2014does}
D.~Lazar, H.~Chen, X.~Wang, and N.~Zeldovich, ``Why does cryptographic software fail? a case study and open problems,'' in \emph{Proceedings of 5th Asia-Pacific Workshop on Systems}, 2014, pp. 1--7.

\bibitem{hazhirpasand2020java}
M.~Hazhirpasand, M.~Ghafari, and O.~Nierstrasz, ``Java cryptography uses in the wild,'' in \emph{Proceedings of the 14th ACM/IEEE International Symposium on Empirical Software Engineering and Measurement (ESEM)}, 2020, pp. 1--6.

\end{thebibliography}

\end{document}